\documentclass{jfm}

\usepackage{graphicx}
\usepackage{natbib}

\usepackage{amssymb}
\usepackage{amsmath}
\usepackage{subfig}

\ifCUPmtlplainloaded \else
  \checkfont{eurm10}
  \iffontfound
    \IfFileExists{upmath.sty}
      {\typeout{^^JFound AMS Euler Roman fonts on the system,
                   using the 'upmath' package.^^J}%
       \usepackage{upmath}}
      {\typeout{^^JFound AMS Euler Roman fonts on the system, but you
                   dont seem to have the}%
       \typeout{'upmath' package installed. JFM.cls can take advantage
                 of these fonts,^^Jif you use 'upmath' package.^^J}%
      }
  \else
  \fi
\fi

\ifCUPmtlplainloaded \else
  \checkfont{msam10}
  \iffontfound
    \IfFileExists{amssymb.sty}
      {\typeout{^^JFound AMS Symbol fonts on the system, using the
                'amssymb' package.^^J}%
       \usepackage{amssymb}%

      }{}
  \fi
\fi

\ifCUPmtlplainloaded \else
  \IfFileExists{amsbsy.sty}
    {\typeout{^^JFound the 'amsbsy' package on the system, using it.^^J}%
     \usepackage{amsbsy}}
    {\providecommand\boldsymbol[1]{\mbox{\boldmath $##1$}}}
\fi

\providecommand\bnabla{\boldsymbol{\nabla}}
\providecommand\bcdot{\boldsymbol{\cdot}}

\newsavebox{\astrutbox}
\sbox{\astrutbox}{\rule[-5pt]{0pt}{20pt}}

\newcommand\p{\ensuremath{\partial}}

\title[Non-singular BIM equations]{Non-singular boundary integral methods for fluid mechanics applications}

\author[E. Klaseboer, Q. Sun and D. Y. C. Chan]%
{E\ls V\ls E\ls R\ls T\ns K\ls L\ls A\ls S\ls E\ls B\ls O\ls E\ls R$^1$%
  \thanks{Email address for correspondence: evert@ihpc.a-star.edu.sg},\ns
Q\ls I\ls A\ls N\ls G\ns S\ls U\ls N$^2$%
 \thanks{Email address for correspondence: qiang.sun@hotmail.com},\ns\break
\and D\ls E\ls R\ls E\ls K\ns Y.\ls C.\ls C\ls H\ls A\ls N$^{3,4}$}

\affiliation{$^1$Institute of High Performance Computing, 1 Fusionopolis Way, 138632, Singapore\\[\affilskip]
$^2$Department of Mechanical Engineering, National University of Singapore, 10 Kent Ridge Crescent, 119260, Singapore\\[\affilskip]
$^3$Department of Mathematics and Statistics, The University of Melbourne, Parkville 3010 VIC Australia\\[\affilskip]
$^4$Faculty Life and Social Sciences, Swinburne University of Technology, Hawthorn 3122 VIC Australia}

\pubyear{2012}
\volume{696}
\pagerange{468--478}
\date{Recieved 29 September 2011; revised 4 January 2012; accepted 3 February 2012.}
\begin{document}

\maketitle

\begin{abstract}
A formulation of the boundary integral method for solving partial differential equations has been developed whereby the usual weakly singular integral and the Cauchy principal value integral can be removed analytically. The broad applicability of the approach is illustrated with a number of problems of practical interest to fluid and continuum mechanics including the solution of the Laplace equation for potential flow, the Helmholtz equation as well as the equations for Stokes flow and linear elasticity.
\end{abstract}


\section{Introduction}\label{sec:intro}
The boundary integral formulation is an efficient method of representing the solutions of certain linear partial differential equations by reducing the dimensionality of the problem by one. The solution in a volume or area domain is represented in terms of an integral over surface(s) or line(s) that enclose the domain. Solutions to fluid dynamics problems that can be modelled by the Laplace equation for potential flow or the Stokes equation for low-Reynolds-number flow as well as continuum mechanics problems such as the Helmholtz equation in scattering problems or problems in linear elasticity can all be represented in terms of such boundary integrals \citep {Becker92}. \cite{Symm1963} provided a practical way to solve the integral equations by treating the surfaces or lines as discrete elements. Since the 1970s the boundary integral method (BIM) has gained increasing prominence \citep[see][for a historical overview]{Chengcheng05}. The advantage of the BIM is self-evident. The reduction in the dimensionality of the problem from a volume (surface) mesh to a surface (line) mesh provides a substantial gain in computational efficiency. However, this gain is offset by the fact that the numerical implementation of the BIM is not straightforward because the approach is plagued by `\textit{a mathematical monster that leaps out of every page}' \citep {Becker92}. In essence, the boundary element formulation uses the Green's function that has a $1/r$ divergence and a $1/r^2$ divergence in its derivative around the source point. The integral over the $1/r$ divergence gives rise to a weak singularity that can be evaluated using semi-analytical techniques. The term from the $1/r^2$ divergence gives rise to a Cauchy principal value (PV) integral that requires careful numerical treatment.

In this communication, we develop a general non-singular boundary integral formulation that is applicable to the Laplace equation for the potential problem, the Helmholtz equation, and equations associated with Stokes flow and linear elastic deformations. The approach is based on removing the singularities in the BIM formulation by subtracting the solution of a special related problem. We demonstrate the details of our approach using the potential problem from which it is easy to see how the method can be extended to the more complicated cases of Stokes flow and linearly elastic deformations. Validation of the approach is obtained by comparing numerical results for problems in Stokes flow for which analytic results are known. This non-singular boundary integral formulation simplifies numerical solutions based on this popular technique.

\begin{figure}
\centering
\subfloat[]{{\label{fig:x0inD}}\includegraphics[width=0.42\textwidth]{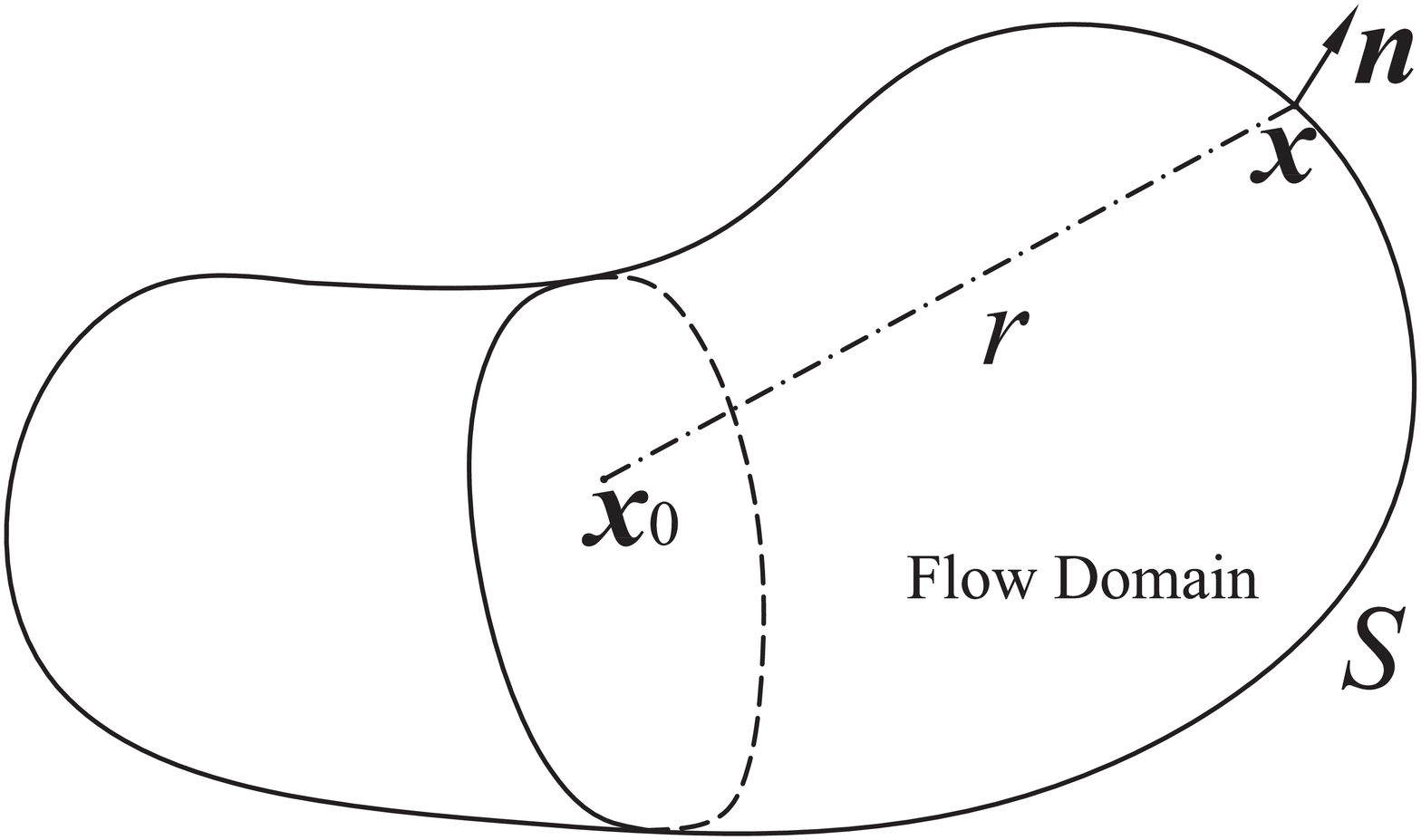}}\hspace{1cm}
\subfloat[]{{\label{fig:x0ionS}}\includegraphics[width=0.42\textwidth]{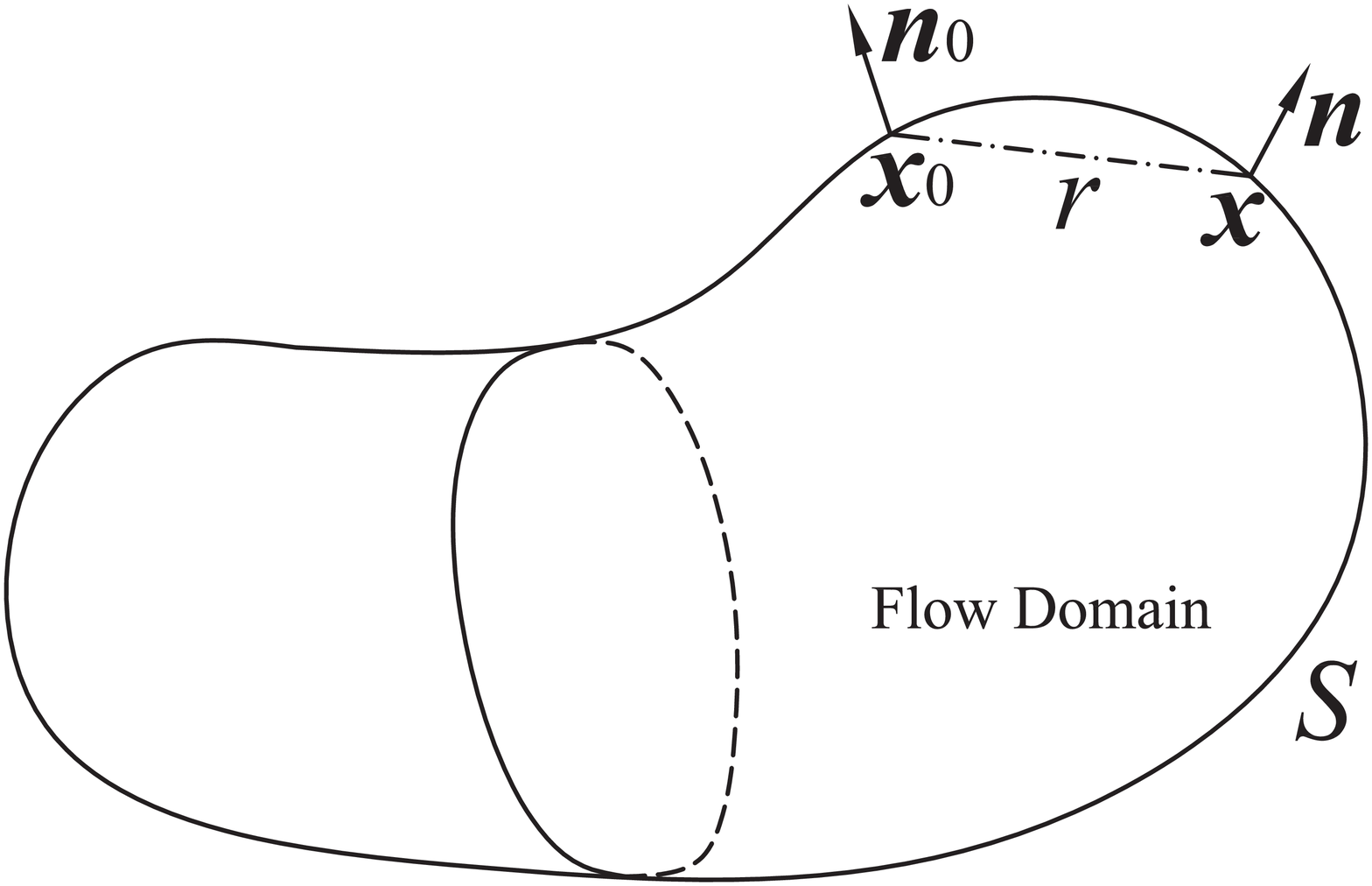}}
\caption{Illustration of the BIM applied to flow problems in an arbitrary three-dimensional flow domain with closed surface $S$: (a) $\boldsymbol{x}_0$ is inside the flow domain; (b) $\boldsymbol{x}_0$ is on the flow domain surface $S$.}\label{fig:illus}
\end{figure}
\section{Potential problem --- non-singular boundary integral formulation }\label{sec:potentialflow}
In fluid dynamics, the potential problem arises in incompressible, inviscid or high-Reynolds-number flows. The velocity field $\boldsymbol{u}=\bnabla\phi$ can be expressed in terms of a scalar potential that satisfies the Laplace equation $\bnabla^{2}\phi=0$. The corresponding free space Green's function $G(\boldsymbol{x},\boldsymbol{x}_0)=1/r$, $r=|\boldsymbol{x}-\boldsymbol{x}_0|$ satisfies $\bnabla^{2}G = -4\pi\delta(\boldsymbol{x}-\boldsymbol{x}_0)$, where $\delta(\boldsymbol{x}-\boldsymbol{x}_0)$ is the Dirac $\delta$-function. Consider the solution in the fluid domain enclosed by the surface $S$. With the help of Green's second identity \citep {Becker92}, the solution at $\boldsymbol{x}_0$ \textit{inside} the domain can be written as the surface integral (see Figure \ref{fig:x0inD})
\begin{align}\label{eq:ptflbie}
4\pi\phi(\boldsymbol{x}_0)+\int_{S}\phi(\boldsymbol{x})\bnabla G(\boldsymbol{x}, \boldsymbol{x}_0)\bcdot \boldsymbol{n}\text{ d}S(\boldsymbol{x}) = \int_{S} G(\boldsymbol{x},\boldsymbol{x}_0)\bnabla \phi(\boldsymbol{x})\bcdot \boldsymbol{n}\text{ d}S(\boldsymbol{x}).
\end{align}

This integral relates the potential $\phi$ at $\boldsymbol{x}_0$ inside the domain to integrals over $\phi$ and its normal derivative $\bnabla\phi\bcdot\boldsymbol{n}=\p{\phi}/\p{n}$, on the surface $S$ with $\boldsymbol{n}$ being the outward unit normal. The vector $\boldsymbol{x}$ points to the integration position on the surface $S$. By letting $\boldsymbol{x}_0$ onto the surface $S$, we have an equation that can be solved for $\phi$ (or $\p{\phi}/\p{n}$) on the surface if $\p{\phi}/\p{n}$ (or $\phi$) is specified. This corresponds to the Dirichlet (or Neumann) problem. This is the boundary integral formulation. However, with $\boldsymbol{x}_0$ on the surface, then as $\boldsymbol{x}\rightarrow\boldsymbol{x}_0$, the integral involving $G$ with a $1/r$ singularity (the single layer term) has a weak singularity that can be handled numerically by changing to local polar coordinates on the surface whereas the integral involving $\bnabla G$ with a $1/r^2$ singularity (the double layer term) gives rise to a Cauchy principal value (PV) integral and a Dirac $\delta$-function contribution. Thus with $\boldsymbol{x}_0$ on the surface $S$ as illustrated in Figure \ref{fig:x0ionS}, the boundary integral equation that needs to be solved is
\begin{align}\label{eq:ptflbie2}
(4\pi-c)\phi(\boldsymbol{x}_0)+\int_{S,\text{PV}}\phi(\boldsymbol{x})\bnabla G(\boldsymbol{x}, \boldsymbol{x}_0)\bcdot \boldsymbol{n}\text{ d}S(\boldsymbol{x}) = \int_{S} G(\boldsymbol{x},\boldsymbol{x}_0)\bnabla \phi(\boldsymbol{x})\bcdot \boldsymbol{n}\text{ d}S(\boldsymbol{x}),
\end{align}
where $c$ is the solid angle subtended at $\boldsymbol{x}_0$ with $c = 2\pi$ if the surface has a defined curvature at $\boldsymbol{x}_0$. The numerical evaluation of the weak singularity associated with $G$ and the Cauchy principal value integral associated with $\bnabla G$ in Eq. (\ref{eq:ptflbie2}) requires special considerations as ordinary integration methods such as Gaussian quadrature can no longer be used (Becker 1992). Different numerical methods have been developed to handle these singularities (see for example \cite{Lean85}, \cite{Bazhlekov2004}). Thus, if either the potential or the normal velocity, $\p{\phi}/\p{n}$, is known on the surface $S$, the other unknown quantity can be calculated (see for example \cite {Gonzalez2011}, \cite {Fong2009}, \cite {Blake1986}, \cite {Wrobel2002}, \cite {Wang1998} and \cite {Zhang2001}). The aim therefore is to avoid the numerical effort needed when having to deal with these singularities.

We recapitulate the earlier work of \citet{Klaseboer2009} and show that both singular terms associated with $G$ and $\bnabla G$ in Eq. (\ref{eq:ptflbie2}) can be removed by considering the linear potential function
\begin{align}\label{eq:ptflpsiex}
\psi(\boldsymbol{x}) = \phi(\boldsymbol{x}_0)+\boldsymbol{a}(\boldsymbol{x}_0)\bcdot(\boldsymbol{x}-\boldsymbol{x}_0),
\end{align}
where the vector $\boldsymbol{a}(\boldsymbol{x}_0)$ will be chosen to eliminate the singularities in Eq. (\ref{eq:ptflbie2}). Clearly $\psi(\boldsymbol{x})$ satisfies $\bnabla^{2}\psi=0$ and the Green's identity, Eq. (\ref{eq:ptflbie}):
\begin{align}\label{eq:eq4}
4\pi\psi(\boldsymbol{x}_0)+\int_{S}\psi(\boldsymbol{x})\bnabla G(\boldsymbol{x}, \boldsymbol{x}_0)\bcdot \boldsymbol{n}\text{ d}S(\boldsymbol{x}) = \int_{S} G(\boldsymbol{x},\boldsymbol{x}_0)\bnabla \psi(\boldsymbol{x})\bcdot \boldsymbol{n}\text{ d}S(\boldsymbol{x}),
\end{align}
with $\boldsymbol{x}_0$ inside the domain. Thus subtracting Eq. (\ref{eq:eq4}) from Eq. (\ref{eq:ptflbie}) and using Eq. (\ref{eq:ptflpsiex}) gives
\begin{align}\label{eq:eq5}
\int_{S}[\phi(\boldsymbol{x})-\psi(\boldsymbol{x})]\bnabla G(\boldsymbol{x}, \boldsymbol{x}_0)\bcdot \boldsymbol{n}\text{ d}S(\boldsymbol{x}) = \int_{S} G(\boldsymbol{x},\boldsymbol{x}_0)\bnabla [\phi(\boldsymbol{x})-\psi(\boldsymbol{x})]\bcdot \boldsymbol{n}\text{ d}S(\boldsymbol{x}).
\end{align}
When $\boldsymbol{x}_0$ is located on the surface $S$, the singularities in Eq. (\ref{eq:eq5}) can be eliminated with the following choice of the vector
\begin{align}\label{eq:eq6}
\boldsymbol{a}(\boldsymbol{x}_0)=[\bnabla \phi(\boldsymbol{x}_0)\bcdot\boldsymbol{n}_0] \boldsymbol{n}_0 \equiv \left(\frac{\p{\phi}}{\p{n}}\right)_0 \hspace{1 mm} \boldsymbol{n}_0,
\end{align}
that depends on the point $\boldsymbol{x}_0$, where the outward unit normal is $\boldsymbol{n}_0 \equiv \boldsymbol{n}(\boldsymbol{x}_0)$ (see Figure \ref{fig:x0ionS}). Finally the required result of the non-singular formulation of the boundary integral equation, with $\boldsymbol{x}_0$ \textit{now located on the surface S}, takes the form
\begin{align}\label{eq:eq7}
&\int_{S}\left[\phi(\boldsymbol{x})-\phi(\boldsymbol{x}_0)-\left(\frac{\p{\phi}}{\p{n}}\right)_0 \boldsymbol{n}_0\bcdot(\boldsymbol{x}-\boldsymbol{x}_0)\right]\bnabla G(\boldsymbol{x}, \boldsymbol{x}_0)\bcdot \boldsymbol{n}\text{ d}S(\boldsymbol{x})\nonumber\\ = &\int_{S} G(\boldsymbol{x},\boldsymbol{x}_0)\left[\frac{\p{\phi}}{\p{n}}-\left(\frac{\p{\phi}}{\p{n}}\right)_0\boldsymbol{n}_0\bcdot \boldsymbol{n}\right]\text{ d}S(\boldsymbol{x}).
\end{align}
This result supersedes the traditional form of the boundary integral formulation given in Eq. (\ref{eq:ptflbie2}) because all singularities have now been removed. In particular, as $\boldsymbol{x}\rightarrow\boldsymbol{x}_0$ which is now on the surface $S$, the weak singularity associated with the integral over $G$ has been eliminated because $[(\p\phi/\p n)-(\p\phi/\p n)_0 \boldsymbol{n}_0\bcdot\boldsymbol{n}]\rightarrow 0$. In the integral over $\bnabla G$, there will no longer be a Cauchy principal value integral or Dirac $\delta$-function contribution as in Eq. (\ref{eq:ptflbie2}). The approach in Eq. (\ref{eq:eq7}) that we present here will give a relationship between $\phi$ and ${\p{\phi}}/{\p{n}}$ as in the original problem. The numerical algorithm for solving Eq. (\ref{eq:eq7}) using the BIM is given by \citet{Klaseboer2009}. A proof of the convergence of the integrals in Eq. (\ref{eq:eq7}) is given in the Appendix. \citet{LiuRudolphi1999} have suggested a similar method of removing the singularities via a Taylor series expansion. Unfortunately, when their approach is implemented, it would give a relationship between $\phi$, ${\p{\phi}}/{\p{n}}$ and $\bnabla \phi(\boldsymbol{x}_0)\bcdot \boldsymbol{n}$, which requires knowledge of $\bnabla \phi(\boldsymbol{x}_0)\bcdot \boldsymbol{n}$.

The surface integral in Eq. (\ref{eq:eq7}) is taken over all surfaces that enclose the domain. In particular, for problems in an infinite domain outside the surface $S$, one must also take into account the `surface at infinity' which will give an additional term $4\pi\phi(\boldsymbol{x}_0)$ on the left hand-side of Eq. (\ref{eq:eq7}), see \cite{Klaseboer2009}, \cite{LiuRudolphi1991} and \cite{LiuRudolphi1999}.

\section{Helmholtz problem --- non-singular boundary integral formulation}\label{sec:helmholtz}
For the solution of the Helmholtz equation $\bnabla^{2}\phi+k^2\phi=0$, in which $k$ is the wave number, we have the free space Green's function $H=\cos(kr)/r$, $r=|\boldsymbol{x}-\boldsymbol{x}_0|$ that satisfies
$\bnabla^{2}H+k^2H=-4\pi\delta(\boldsymbol{x}-\boldsymbol{x}_0)$. In the same way as we obtained Eq. (\ref{eq:eq5}) for the potential problem, we find, for $\boldsymbol{x}_0$ in the domain
\begin{align}\label{eq:helmeq1}
\int_{S}[\phi(\boldsymbol{x})-\psi(\boldsymbol{x})]\bnabla H(\boldsymbol{x}, \boldsymbol{x}_0)\bcdot \boldsymbol{n}\text{ d}S(\boldsymbol{x}) = \int_{S} H(\boldsymbol{x},\boldsymbol{x}_0)\bnabla [\phi(\boldsymbol{x})-\psi(\boldsymbol{x})]\bcdot \boldsymbol{n}\text{ d}S(\boldsymbol{x}),
\end{align}
where
\begin{align}\label{eq:helmeq2}
\psi(\boldsymbol{x}) = \phi(\boldsymbol{x}_0)\cos{[k\boldsymbol{n}_0\bcdot(\boldsymbol{x}-\boldsymbol{x}_0)}] + \frac{b(\boldsymbol{x}_0)}{k} \sin{[k\boldsymbol{n}_0\bcdot(\boldsymbol{x}-\boldsymbol{x}_0)]}.
\end{align}
Upon putting $\boldsymbol{x}_0$ onto the surface $S$ in Eq. (\ref{eq:helmeq1}), we can eliminate all singular terms with the choice
\begin{align}\label{eq:helmeqb}
b(\boldsymbol{x}_0)=\bnabla \phi(\boldsymbol{x}_0)\bcdot\boldsymbol{n}_0 \equiv \left(\frac{\p{\phi}}{\p{n}}\right)_0.
\end{align}
Thus the non-singular formulation of boundary integral equation for the Helmholtz problem when $\boldsymbol{x}_0$ \textit{is now located on the surface S}, with $\chi\equiv k\boldsymbol{n}_0\bcdot(\boldsymbol{x}-\boldsymbol{x}_0)$, is
\begin{align}\label{eq:hfbiefr}
&\int_{S} \left\{\phi(\boldsymbol{x})-\phi(\boldsymbol{x}_0)\cos\chi - \frac{1}{k}\left(\frac{\p{\phi}}{\p{n}}\right)_0 \sin\chi\right\} \bnabla H(\boldsymbol{x},\boldsymbol{x}_0)\bcdot\boldsymbol{n}\text{d}S(\boldsymbol{x})\nonumber\\
=&\mspace{-8.0mu} \int_{S} \mspace{-8.0mu} H(\boldsymbol{x},\boldsymbol{x}_0)\left\{\frac{\p{\phi}}{\p{n}} + k\phi(\boldsymbol{x}_0)\boldsymbol{n}_0\bcdot\boldsymbol{n}\sin\chi - \left(\frac{\p{\phi}}{\p{n}}\right)_0 \boldsymbol{n}_0\bcdot\boldsymbol{n}\cos\chi\right\}\text{d}S(\boldsymbol{x}).
\end{align}
This is the key result in which all singularities associated with the boundary integral formulation of the Helmholtz problem have been removed. As $\boldsymbol{x}\rightarrow\boldsymbol{x}_0$ , the analytic structure of the integrands Eq. (\ref{eq:hfbiefr}) is essentially the same as that in Eq. (\ref{eq:eq7}) where the integrands do not diverge. In the limit $k\rightarrow0$, this reduces to result Eq.(\ref{eq:eq7}) for the potential problem.

\section{Stokes problem --- non-singular boundary integral formulation}\label{sec:stokesflow}
The governing equations for the pressure, $p$, and velocity field, $\boldsymbol{u}$ for incompressible Stokes flow in a Newtonian fluid of dynamic viscosity $\mu$ are
\begin{align}
-\bnabla p+\mu\bnabla^2\boldsymbol{u}= 0 \hspace{5 mm} \text{and} \hspace{5 mm}\bnabla\bcdot\boldsymbol{u}=0.
\end{align}
The stress tensor $\sigma_{ik}$ is given by
\begin{align}
\mathsfbi{\sigma}_{ik} = -p\mathsfbi{\delta}_{ik}+\mu\left[\frac{\p{u_i}}{\p{x_k}}+\frac{\p{u_k}}{\p{x_i}}\right],
\end{align}
where $\delta_{ik}$ is the Kronecker delta function. Using the Lorentz reciprocal theorem \citep{Lorentz1907}, the velocity component in the $j$-th direction, $u_j^0\equiv u_j(\boldsymbol{x}_0)$, at position $\boldsymbol{x}_0$ in the fluid domain can be written as a boundary integral over the enclosing surface $S$ \citep {Pozrikidis92} as
\begin{eqnarray}\label{eq:stkflbie}
 8\pi u^{0}_{j}+\int_{S} u_i\mathsfbi{T}_{ijk}n_{k}\text{ d}S &=& \frac{1}{\mu}\int_{S}\mathsfbi{\sigma}_{ik} n_{k}\mathsfbi{U}_{ij} \text{ d}S       \nonumber \\
   &=& \frac{1}{\mu}\int_{S}f_{i}\mathsfbi{U}_{ij}\text{ d}S.
\end{eqnarray}
The $i^{th}$ component of the traction vector $\boldsymbol{f}$, is defined as $f_{i}=\mathsfbi{\sigma}_{ik} n_k$. Equation (\ref{eq:stkflbie}) for the Stokes problem is the analogue of (\ref{eq:ptflbie}) for the potential problem. The fundamental solutions for Stokes flow $\mathsfbi{U}_{ij}$ and $\mathsfbi{T}_{ijk}$ are given by \citep {Pozrikidis92}:
\begin{align}
\mathsfbi{U}_{ij}(\boldsymbol{x},\boldsymbol{x}_0) = \frac{\mathsfbi{\delta}_{ij}}{r}+\frac{\skew3\hat{x}_{i}\skew3\hat{x}_{j}}{r^3},
\end{align}
\begin{align}
\mathsfbi{T}_{ijk}(\boldsymbol{x},\boldsymbol{x}_0) = -6\frac{\skew3\hat{x}_i\skew3\hat{x}_j\skew3\hat{x}_k}{r^5},
\end{align}
where $\skew3\hat{x}_i$ etc, are the components of $\skew3\hat{\boldsymbol{x}} = \boldsymbol{x}-\boldsymbol{x}_0$, $r=|\skew3\hat{\boldsymbol{x}}|$ and $n_{k}$ is the $k$-th component of the unit normal of the surface pointing out of the flow domain. The functions $\mathsfbi{U}_{ij}$ and $\mathsfbi{T}_{ijk}$ diverge as $1/r$ and $1/r^2$, respectively, with the same behaviour as $G$ and $\bnabla G$ for the potential problem and give rise to singular behaviour in Eq. (\ref{eq:stkflbie}) when $\boldsymbol{x}_0$ is on the surface $S$. The traditional boundary integral formulation is obtained by putting $\boldsymbol{x}_0$ onto the surface $S$ in Eq. (\ref{eq:stkflbie}), and as in Eq. (\ref{eq:ptflbie2}), this will give rise to Cauchy principal value integrals and Dirac $\delta$-function contributions on the left hand-side and weakly singular integrands on the right hand-side of Eq. (\ref{eq:ptflbie2}). To remove such singularities, consider a zero-pressure linear velocity field
\begin{align}\label{eq:stkfllnvl}
\boldsymbol{w}(\boldsymbol{x}) = \boldsymbol{u}(\boldsymbol{x}_0)+\frac{1}{\mu}\mathsfbi{\boldsymbol{M}}(\boldsymbol{x}_0)\bcdot(\boldsymbol{x}-\boldsymbol{x}_{0}),
\end{align}
where the matrix $\mathsfbi{\boldsymbol{M}}(\boldsymbol{x}_0)$ will be chosen to cancel the arising singularities in Eq. (\ref{eq:stkflbie}) when $\boldsymbol{x}_0$ is on the surface $S$. The symmetric stress tensor corresponding to this linear flow field is
\begin{align}\label{eq:stsigma}
\boldsymbol{\Sigma}(\boldsymbol{x}_0)=\boldsymbol{M}(\boldsymbol{x}_0)+\boldsymbol{M}^T(\boldsymbol{x}_0)
\hspace{5 mm} \text { or } \hspace{5 mm}
\mathsfbi{\Sigma}^{0}_{ik} = \mathsfbi{M}^{0}_{ik}+\mathsfbi{M}^{0}_{ki}.
\end{align}
For the velocity field $\boldsymbol{w}$ to meet the incompressibility condition: $\bnabla \bcdot \boldsymbol{w}(\boldsymbol{x}) = 0$, \begin{align}\label{eq:incomcond}
\text{Tr}[\mathsfbi{\boldsymbol{M}}(\boldsymbol{x}_0)] = \text{Tr}[ \mathsfbi{\boldsymbol{\Sigma}}(\boldsymbol{x}_0) ]/2= 0
\end{align}
must hold. Since $\boldsymbol{w}$ satisfies the equations for Stokes flow, the difference $(\boldsymbol{u - w})$ also satisfies Eq. (\ref{eq:stkflbie}). In component form the difference becomes ($x^0_j = j^{th}$ component of $\boldsymbol{x}_0$)
\begin{align}\label{eq:stkflbiefr}
\int_{S} \left[u_i-u_{i}^{0}-\frac{1}{\mu}\mathsfbi{M}^{0}_{il}(x_{l}-x_{l}^{0})\right]\mathsfbi{T}_{ijk}n_{k}\text{ d}S = \frac{1}{\mu}\int_{S}\left(f_i-\mathsfbi{\Sigma}^{0}_{il}n_{l}\right)\mathsfbi{U}_{ij}\text{ d}S.
\end{align}
This integral equation, with $\boldsymbol{x}_0$ \textit{located on the surface S}, will have no singular behaviour if we choose (adopting the convention of implicit summation over repeated indices)
\begin{align}\label{eq:mmatrix2}
\boldsymbol{M}(\boldsymbol{x}_0)=\boldsymbol{f}(\boldsymbol{x}_0)\boldsymbol{n}(\boldsymbol{x}_0)
-\frac{1}{4}\boldsymbol{f}(\boldsymbol{x}_0) \bcdot \boldsymbol{n}(\boldsymbol{x}_0) \left[\boldsymbol{I}+\boldsymbol{n}(\boldsymbol{x}_0)\boldsymbol{n}(\boldsymbol{x}_0)\right],
\end{align}
\begin{align}\label{eq:mmatrix}
\mathsfbi{M}^{0}_{il} = f_{i}^{0}n^{0}_{l}-\frac{1}{4}(f_{k}^{0}n^{0}_{k})(\mathsfbi{\delta}_{il}+n_{i}^{0}n_{l}^{0}),
\end{align}
and
\begin{align}\label{eq:sigma2}
\boldsymbol{\Sigma}(\boldsymbol{x}_0)=\boldsymbol{f}(\boldsymbol{x}_0)\boldsymbol{n}(\boldsymbol{x}_0)
+\boldsymbol{n}(\boldsymbol{x}_0)\boldsymbol{f}(\boldsymbol{x}_0)
-\frac{1}{2}\boldsymbol{f}(\boldsymbol{x}_0) \bcdot \boldsymbol{n}(\boldsymbol{x}_0) \left[\boldsymbol{I}+\boldsymbol{n}(\boldsymbol{x}_0)\boldsymbol{n}(\boldsymbol{x}_0)\right],
\end{align}
\begin{align}\label{eq:stkfllnvarpsi}
\mathsfbi{\Sigma}^{0}_{il}=\mathsfbi{M}^{0}_{il}+\mathsfbi{M}^{0}_{li} = (f_{i}^{0}n^{0}_{l}+f^{0}_{l}n_{i}^{0})-\frac{1}{2}(f_{k}^{0}n^{0}_{k})(\mathsfbi{\delta}_{il}+n_{i}^{0}n_{l}^{0}).
\end{align}
The relation between the stress tensor $\boldsymbol{\Sigma}(\boldsymbol{x}_0)$ and the matrix $\boldsymbol{M}(\boldsymbol{x}_0)$ in terms of the traction $\boldsymbol{f}(\boldsymbol{x}_0)$ and the surface normal $\boldsymbol{n}(\boldsymbol{x}_0)$ needed to ensure Eq.(\ref{eq:stkflbiefr}) is non-singular is in fact not unique. It is easy to verify that the expressions in Eqs. (\ref{eq:mmatrix2}) to (\ref{eq:stkfllnvarpsi}) obey those constraints in Eq. (\ref{eq:incomcond}) and as a result, the integrands in Eq.(\ref{eq:stkflbiefr}) are not singular as $\boldsymbol{x}\rightarrow\boldsymbol{x}_0$ on the surface (see the proof in Appendix). Thus, Eqs. (\ref{eq:stkflbiefr}) to (\ref{eq:stkfllnvarpsi}) form the non-singular boundary integral formulation of the Stokes problem. Analogous to the potential problem, the elements of the matrix $\boldsymbol{M}(\boldsymbol{x}_0)$ vary for each $\boldsymbol{x}_0$ on the surface $S$.

\section{Linear elasticity problem --- non-singular boundary integral formulation}\label{sec:linearelastic}
The regularisation method described thus far is quite general. The non-singular boundary integral formulation of the Stokes problem can be adapted to the linear elastic problem in solid mechanics as follows. The strain tensor $\mathsfbi{\varepsilon}$ in the elastic problem is defined in terms of the position vector field $\boldsymbol{u}$
\begin{align}
\mathsfbi{\varepsilon}_{ij}=\frac{1}{2}\left(\frac{\p{u_i}}{\p{x_j}}+\frac{\p{u_j}}{\p{x_i}}\right).
\end{align}
For a linear elastic material in equilibrium and in the absence of body forces, the stress tensor satisfies $\p{{\sigma}_{ij}}/\p{x_j}=0$ and is given by
\begin{align}
\mathsfbi{\sigma}_{ij}=\frac{2\mathsfbi{\mu \nu}}{1-2\nu}\mathsfbi{\delta}_{ij}\mathsfbi{\varepsilon}_{ll}+2\mu\mathsfbi{\varepsilon}_{ij},
\end{align}
where the Poisson ratio $\nu$, and the shear modulus $\mu$ are related to the Young's modulus, $E = 2 \mu (1+\nu)$. The displacement field $\boldsymbol{u}(\boldsymbol{x}_0)$ at an interior point of the elastic material can be expressed in terms of integrals over the displacement field and the surface traction on the enclosing surface $S$ by the same equation as Eq. (\ref{eq:stkflbie}), except the fundamental solutions $\mathsfbi{U}_{ij}$ and $\mathsfbi{T}_{ijk}$ for the linear elastic problem are now given by \citet{Becker92}:
\begin{align}
\mathsfbi{U}_{ij} = \frac{1}{2(1-\nu)}\left[ (3-4\nu)\frac{\mathsfbi{\delta}_{ij}}{r}+\frac{\skew3\hat{x}_{i}\skew3\hat{x}_{j}}{r^3}\right],
\end{align}
and
%\begin{align}
%\mathsfbi{T}_{ijk}= -\frac{1}{1-\nu}\left[(1-2\nu)\mathsfbi{\delta}_{ij}\frac{\skew3\hat{x}_{k}}{r^3} +3\frac{\skew3\hat{x}_i\skew3\hat{x}_j\skew3\hat{x}_k}{r^5}\right]- \frac{1-2\nu}{1-\nu}\left[\frac{\skew3\hat{x}_{j}}{r^3}n_in_k-\frac{\skew3\hat{x}_{i}}{r^3}n_jn_k\right].
%\end{align}
\begin{align}
\mathsfbi{T}_{ijk}= -\frac{1}{1-\nu}  \left[  3\frac{\skew3\hat{x}_i\skew3\hat{x}_j\skew3\hat{x}_k}{r^5}  + \frac{1-2\nu} {r^3} \left( - \delta_{ij}\hat{x}_k + \delta_{jk}\hat{x}_i + \delta_{ki}\hat{x}_j \right)   \right] .
\end{align}
With these replacements, Eqs. (\ref{eq:stkflbiefr}) to (\ref{eq:stkfllnvarpsi}) are also the non-singular boundary integral formulation of the linear elastic problem for the displacement field $\boldsymbol{u}$ with the traction vector $\boldsymbol{f}$ defined by $f_i=\mathsfbi{\sigma}_{ik}n_{k}$. For an incompressible material: $\nu=1/2$, then $\mathsfbi{U}_{ij}$ and $\mathsfbi{T}_{ijk}$ for the linear elastic problem and the Stokes problem become identical.

\begin{figure}
\centering
\includegraphics[width=0.75\textwidth]{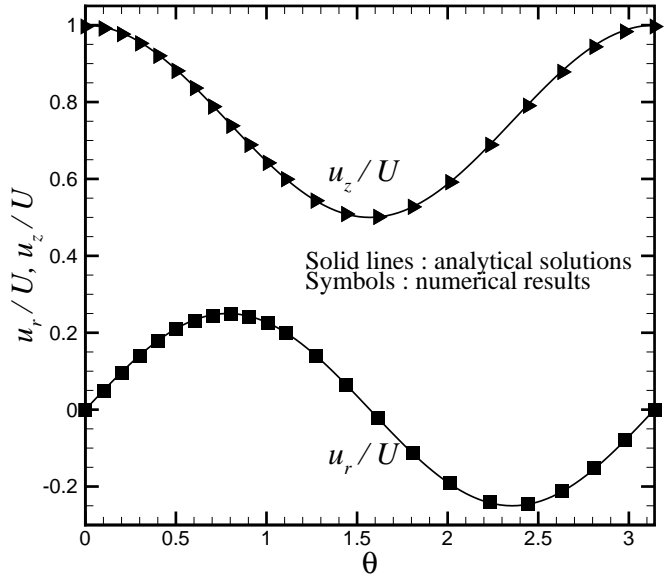}
\caption{Distributions of the normalized velocities $u_r/U$ and $u_z/U$ along $\theta$ in the second test case for Stokes problem of a spherical bubble rising under buoyancy force.}\label{fig:numrslts}
\end{figure}

\section{Discussion and implementation}\label{sec:conclusion}
In this communication we have developed a non-singular boundary integral formulation for solving four common and related problems in hydrodynamics and solid mechanics. The common theme in the formulation is the removal of the singularities associated with the traditional boundary integral formulation by subtracting a simpler solution of a related problem with an appropriate choice of the free parameter in the solution.

The numerical implementation of our non-singular formulation for the potential problem, Eq. (\ref{eq:eq7}), has been described in \citet{Klaseboer2009}. The Stokes problem is very similar, except that the matrix elements appear in blocks of sub-matrices. When the surface is discretised in nodes and elements, the usual Gaussian-quadrature integration procedure can be applied for all elements, including the singular ones. This will result in a system of equations relating the potential and its normal derivative through two influence matrices, which is the discretized equivalent of Eq. (\ref{eq:eq7}). The previously singular contributions can be found on the diagonals of the influence matrices, one corresponding to $G(\boldsymbol{x}, \boldsymbol{x}_0)$ and one to its normal derivative. All of the terms corresponding to $\boldsymbol{x}_0$ now correspond to those contributions and can be obtained by simple summation. Several examples for which analytical solutions exist were tested (see \citet{Klaseboer2009} for more details).

For the Stokes problem, the singularities appear in blocks of $3\times3$ around the diagonals of the influence matrices. A procedure very similar to that followed for the potential flow can be followed to get those values by summation once more, based on Eq. (\ref{eq:stkflbiefr}).

Two examples are provided for the Stokes flow implementation. Both use flat three-noded linear elements in which the surface representation and the shape functions are linear. The first example is that of a sphere moving with a constant velocity $\boldsymbol{U}$. The mesh is similar to that used in \citet{Klaseboer2009}. Thus the velocity vectors, $\boldsymbol{u}$, are given at all nodes as $\boldsymbol{u} = \boldsymbol{U}$. The traction, $\boldsymbol{f}$, is then calculated and, within the expected discretisation error, agrees with the analytic solution $\boldsymbol{f} = 3\mu\boldsymbol{U}/(2R)$, where $R$ is the radius of the sphere.

In the second test case, the traction $\boldsymbol{f}$ is given instead and the velocity $\boldsymbol{u}$ is calculated. We use the same test case as presented in \citet{Pigeonneau2011}, and take $\boldsymbol{f} = -\rho gz\boldsymbol{n}$ that corresponds to a spherical bubble rising under buoyancy force, where $g$ is the magnitude of gravity, and $\rho$ is the fluid density. The exact solution in cylindrical coordinates is given by Eqs. (34) and (35) in \citet{Pigeonneau2011}: $u_r=U\sin(2\theta)/4$ and $u_z = U(1-\sin^2\theta/2)$, where $U = \rho g R^2/(3\mu)$, and $\theta$ is the angle between the unit vector in the $z$-direction and the radial direction. The results are shown in Figure \ref{fig:numrslts}. Even for a mesh consisting of only $252$ nodes ($500$ elements) the accuracy is within $2$\%.

The present non-singular formulation therefore offers all the advantages associated with the reduction of dimension afforded by the boundary integral technique without the extra numerical effort needed to handle the singularities that arise with the traditional boundary integrals formulation. Although the size of the numerical problem remains the same, the absence of singularities means that there will be a significant reduction in coding effort which will minimise the opportunity for coding error.

We believe the present contribution is a novel advance that will have both pedagogical and practical implications.

\begin{acknowledgements}
EK would like to thank A. Prosperetti for stimulating discussions and B. C. Khoo for maintaining his interest in practical applications of boundary element methods. DYCC is a Visiting Scientist at the IHPC and an Adjunct Professor at the National University of Singapore. This work is supported in part by the Australian Research Council Discovery Project Grant Scheme.
\end{acknowledgements}

\oneappendix
\section{Non-singular proof}
We show that the integrands in Eq. (\ref{eq:eq7}) for the potential flow problems and Eq. (\ref{eq:stkflbiefr}) for the Stokes flow problems are non-singular by analysing the analytic behaviour of the integrand in the neighbourhood of $\boldsymbol{x}_0$. Define a Cartesian system $(\xi, \eta, \zeta)$ with $\boldsymbol{x}_0=(0,0,0)$ as the origin and $\boldsymbol{n}_0 = (0,0,1)$. In the neighbourhood of $\boldsymbol{x}_0$, a point $\boldsymbol{x} = (\xi, \eta, \zeta)$ that lies on the surface $S$ with a suitable choice of the local coordinates, $\xi$, $\eta$ and $\zeta$, must satisfy
\begin{align}
S = \zeta + \frac{1}{2}a_{s} \xi^2+\frac{1}{2}b_{s}\eta^2=0
\end{align}
where higher order terms of $O(\xi^3, \eta^3)$ have been omitted. The constants $a_{s}$ and $b_{s}$ are related to the principal curvatures of $S$ at $\boldsymbol{x}_0$ and $\zeta$ is quadratic in $\xi$ and $\eta$. The unit normal vector at $\boldsymbol{x}$ is $\boldsymbol{n}=\bnabla{S}/|\bnabla{S}|=\cos{\gamma}\text{ }(a_{s}\xi,b_{s}\eta,1)$, where $\cos{\gamma} = [1+(a_{s}\xi)^2+(b_{s}\eta)^2]^{-1/2}$ is the direction cosine. The differential surface is $\text{d}S(\boldsymbol{x}) = \text{d}\xi\text{d}\eta/\cos{\gamma}$. The Green function has the form: $G(\boldsymbol{x},\boldsymbol{x}_0)=(\xi^2+\eta^2+\zeta^2)^{-1/2}$.

With these preliminary results, the integral on the left hand-side of Eq. (\ref{eq:eq7}), has the following limiting form obtained by using a Taylor expansion about $\boldsymbol{x}_0$ in terms of the local coordinates
\begin{align}
&\int\limits_{\boldsymbol{x}\rightarrow\boldsymbol{x}_0} \left[\phi(\boldsymbol{x})-\phi(\boldsymbol{x}_0)-\left(\frac{\p{\phi}}{\p{n}}\right)_0 \boldsymbol{n}_0\bcdot(\boldsymbol{x}-\boldsymbol{x}_0)\right]\bnabla G(\boldsymbol{x}, \boldsymbol{x}_0)\bcdot \boldsymbol{n}\text{ d}S(\boldsymbol{x})\nonumber\\
\sim &\int\limits_{\boldsymbol{x}\rightarrow\boldsymbol{x}_0} \left[\xi\left(\frac{\p{\phi}}{\p{\xi}}\right)_{0} + \eta\left(\frac{\p{\phi}}{\p{\eta}}\right)_{0} \right]\left[\frac{1}{2}a_{s}\xi^2+\frac{1}{2}b_{s}\eta^2\right] \frac{\text{d}\xi\text{d}\eta}{(\xi^2+\eta^2+\zeta^2)^{3/2}}.
\end{align}
We see that both the numerator and the denominator of the integrand are of $O(\xi^3,\eta^3)$, thus the integrand remains finite, as $\xi\rightarrow 0$ and $\eta\rightarrow 0$, that is $\boldsymbol{x}\rightarrow \boldsymbol{x}_0$. Furthermore, if the surface around $\boldsymbol{x}_0$ is a planar element, the constants $a_{s}$ and $b_{s}$ will be zero and the integrand vanishes. For non-planar elements, the point-wise discontinuity at $(\xi, \eta) = (0, 0)$ is a set of measure zero and therefore does not contribute to the value of the integral.

Similarly, as $\boldsymbol{x}\rightarrow \boldsymbol{x}_0$, the integral on the right hand-side of Eq. (\ref{eq:eq7}), has the form
\begin{align}
&\int\limits_{\boldsymbol{x}\rightarrow\boldsymbol{x}_0} \mspace{-8.0mu}G(\boldsymbol{x},\boldsymbol{x}_0)\left[\frac{\p{\phi}}{\p{n}}-\left(\frac{\p{\phi}}{\p{n}}\right)_0\boldsymbol{n}_0\bcdot \boldsymbol{n}\right]\text{ d}S(\boldsymbol{x})\nonumber\\
\sim & \mspace{-8.0mu}\int\limits_{\boldsymbol{x}\rightarrow\boldsymbol{x}_0}\mspace{-8.0mu} \left[(a_{s}\xi)\mspace{-5.0mu}\left(\frac{\p{\phi}}{\p{\xi}}\right)_{0}\mspace{-8.0mu} + (b_{s}\eta)\mspace{-5.0mu}\left(\frac{\p{\phi}}{\p{\eta}}\right)_{0}\mspace{-8.0mu} + \xi\mspace{-5.0mu}\left(\frac{\p^2{\phi}}{\p{\xi}\p{\zeta}}\right)_{0}\mspace{-8.0mu} + \eta\mspace{-5.0mu}\left(\frac{\p^2{\phi}}{\p{\eta}\p{\zeta}}\right)_{0} \right] \frac{\text{d}\xi\text{d}\eta}{(\xi^2+\eta^2+\zeta^2)^{1/2}}.
\end{align}
We see that both the numerator and the denominator of the integrand vanish linearly with $\xi$ and $\eta$, as $\xi\rightarrow 0$ and $\eta\rightarrow 0$. Thus the integrand has no divergences. Again, if the surface around $\boldsymbol{x}_0$ is a planar element and the normal derivative $(\p{\phi}/\p{\zeta})$ is constant over that element, then the integrand vanishes. For non-planar elements, the point-wise discontinuity at $(\xi, \eta) = (0, 0)$ is a set of measure zero and therefore does not contribute to the value of the integral. This has been demonstrated for quadratic elements \citep{Klaseboer2009}.

This completes the proof that Eq. (\ref{eq:eq7}) is non-singular. In the same way, Eq. (\ref{eq:hfbiefr}) for the Helmholtz problem can also be shown to be non-singular.

Using the same local coordinate system, we can also show that Eq. (\ref{eq:stkflbiefr}) for the Stokes problem is not singular. First we consider the integral on the left hand-side of Eq. (\ref{eq:stkflbiefr}). As $\boldsymbol{x}\rightarrow \boldsymbol{x}_0$, it is straightforward to show using a Taylor expansion that the two terms in the integrand have the limiting form using the notation $\skew3\hat{\boldsymbol{x}} \equiv \boldsymbol{x} - \boldsymbol{x}_0 = (\xi, \eta, \zeta)$, $\boldsymbol{f}(\boldsymbol{x}) = (f_1, f_2, f_3)$ and $\boldsymbol{f}(\boldsymbol{x}_0) = (f^0_1, f^0_2, f^0_3)$
\begin{align}
&\int\limits_{\boldsymbol{x}\rightarrow\boldsymbol{x}_0} \mspace{-8.0mu} (u_i-u_{i}^{0})\mathsfbi{T}_{ijk}n_{k}\text{ d}S\nonumber\\ \sim & \mspace{-8.0mu} \int\limits_{\boldsymbol{x}\rightarrow\boldsymbol{x}_0} \mspace{-8.0mu} -\frac{6(a_{s}\xi^2+b_{s}\eta^2+\zeta)}{(\xi^2+\eta^2+\zeta^2)^{5/2}} \Bigg\{\left[\xi\left(\frac{\p{u_i}}{\p{\xi}}\right)_0 + \eta\left(\frac{\p{u_i}}{\p{\eta}}\right)_0 + \zeta\left(\frac{\p{u_i}}{\p{\zeta}}\right)_0\right]\skew3\hat{x}_i\Bigg\}\skew3\hat{x}_j\text{ d}\xi\text{d}\eta,\\
&\int\limits_{\boldsymbol{x}\rightarrow\boldsymbol{x}_0} \mspace{-8.0mu}  \mathsfbi{M}^{0}_{il}\skew3\hat{x}_{l}\mathsfbi{T}_{ijk}n_{k}\text{ d}S\nonumber \\
\sim & \mspace{-8.0mu} \int\limits_{\boldsymbol{x}\rightarrow\boldsymbol{x}_0} \mspace{-8.0mu} -\frac{6(a_{s}\xi^2+b_{s}\eta^2+\zeta)[\xi(f^0_1\zeta-f^0_3\xi/4)+ \eta(f^0_2\zeta-f^0_3\eta/4)+f^0_3\zeta^2/2]}{(\xi^2+\eta^2+\zeta^2)^{5/2}}\text{ }\skew3\hat{x}_j\text{ d}\xi\text{d}\eta.
\end{align}
The numerator and the denominator of both terms are of $O(\xi^5,\eta^5)$ and, thus, they approach constant values as $\boldsymbol{x}\rightarrow \boldsymbol{x}_0$, and as a consequence, the integral on the left hand-side of Eq. (\ref{eq:stkflbiefr}) is non-singular.

Turning now to the integral on the right hand-side of Eq. (\ref{eq:stkflbiefr}) where in the limit $\boldsymbol{x}\rightarrow \boldsymbol{x}_0$, the integrand has the limiting form
\begin{align}
&\int\limits_{\boldsymbol{x}\rightarrow\boldsymbol{x}_0} \mspace{-8.0mu} \left(f_i-\mathsfbi{\Sigma}^{0}_{il}n_{l}\right)\mathsfbi{U}_{ij}\text{ d}S \sim \int\limits_{\boldsymbol{x}\rightarrow\boldsymbol{x}_0} \frac{(1-\cos{\gamma})(f^0_1\xi + f^0_2\eta + f^0_3\zeta) + L_i\skew3\hat{x}_i}{\cos{\gamma}\text{ }(\xi^2+\eta^2+\zeta^2)^{3/2}}\text{ } \skew3\hat{x}_j\text{ d}\xi\text{d}\eta\nonumber\\
+ &\int\limits_{\boldsymbol{x}\rightarrow\boldsymbol{x}_0} \frac{(1-\cos{\gamma})f^{0}_j+L_{j} - \left[\cos{\gamma}\left(f^0_1a_s\xi+f^0_2b_s\eta\right)\right]n^0_j+(f^0_3/2)(n_j-\cos{\gamma}\text{ }n^0_j)} {\cos{\gamma}\text{ }(\xi^2+\eta^2+\zeta^2)^{1/2}}\text{ d}\xi\text{d}\eta\nonumber\\
+&\int\limits_{\boldsymbol{x}\rightarrow\boldsymbol{x}_0} \frac{(f^0_3/2)(a_s\xi^2+b_s\eta^2) - \left(f^0_1a_s\xi+f^0_2b_s\eta\right)\zeta} {(\xi^2+\eta^2+\zeta^2)^{3/2}}\text{ } \skew3\hat{x}_j\text{ d}\xi\text{d}\eta
,
\end{align}
where $L_i \equiv \xi(\p{f_i}/\p{\xi})_0+\eta(\p{f_i}/\p{\eta})_0+\zeta(\p{f_i}/\p{\zeta})_0$. The numerator and the denominator of the first term are of $O(\xi,\eta)$ and those of the second and third terms are of $O(\xi^3,\eta^3)$ and thus all terms approach constant values as $\boldsymbol{x}\rightarrow \boldsymbol{x}_0$, and so it follows that the integral on the right hand-side of Eq. (\ref{eq:stkflbiefr}) is non-singular.

For the case in which the surface $S$ is a planar element for which the curvatures $a_{s}$ and $b_{s}$ are zero, $\cos{\gamma} = 1$ and the traction $\boldsymbol{f}(\boldsymbol{x}_0)$ is constant within the plane, the integrands of both integrals in Eq. (\ref{eq:stkflbiefr}) will vanish. For non-planar elements, the point-wise discontinuity at $(\xi, \eta) = (0, 0)$ is a set of measure zero and therefore does not contribute to the value of the integral.

This completes the proof that Eq. (\ref{eq:stkflbiefr}) is non-singular. In the same way, the linear elasticity problem can also shown to be non-singular.

\bibliographystyle{jfm}

\bibliography{GSRBEM_RefV03}

\end{document}